\documentclass[reprint,amsmath,amssymb,aps,prl,superscriptaddress, showpacs, eprint]{revtex4-2}
\usepackage[utf8]{inputenc}

\usepackage{amsmath,amssymb,ascmac,fancybox, multirow}
\usepackage{color}
\usepackage{graphicx}
\usepackage{url}
\usepackage{siunitx}
\usepackage{bm}
\usepackage{physics}
\usepackage{mathtools}
\usepackage{hyperref}
\usepackage[T1]{fontenc}
\usepackage{ulem}
\usepackage{cleveref}
\graphicspath{figures/}

\newcommand{\ani}[1]{{#1}}
\newcommand{\cre}[1]{{#1}^\dagger}

\newcommand{\ene}{\varepsilon}
\renewcommand{\vec}{\vb*}

\newcommand{\up}{\uparrow}
\newcommand{\down}{\downarrow}

\newcommand{\red}[1]{{#1}}

\begin{document}
\date{\today}
\title{Topological bound on structure factor}%
\author{Yugo Onishi}
\affiliation{Department of Physics, Massachusetts Institute of Technology, Cambridge, MA 02139, USA}
\author{Liang Fu}
\affiliation{Department of Physics, Massachusetts Institute of Technology, Cambridge, MA 02139, USA}

\begin{abstract}
We show that the static structure factor of general many-body systems with $U(1)$ symmetry has a lower bound determined only by the ground state Chern number. Our bound relies only on causality and non-negative energy dissipation, and holds for a wide range of \red{two-dimensional gapped} systems. We apply our theory to (fractional) Chern insulators, (fractional) quantum spin Hall insulators, topological superconductors, and chiral spin liquids. Our results uncover a universal feature of topological phases beyond the quantized response.
\end{abstract}

\maketitle

It is well known that topological invariants of ground states often govern quantized physical responses of quantum many-body systems: for example, the many-body Chern number determines the quantized (anomalous) Hall conductivity~\cite{klitzing_new_1980, Thouless1982, niu_quantized_1985, Chang2013, park_observation_2023, xu_observation_2023, lu_fractional_2023}. %
Recently, it has been shown~\cite{onishi_fundamental_2024} that the ground-state topology not only dictates the quantization of certain observables, but also sets universal bounds on other physical quantities. Specifically, we found that the topology of (fractional) Chern insulators sets an upper bound on the energy gap, which is determined only by the electron density, mass and the many-body Chern number.
This topological bound on the energy gap reveals new information about Chern insulators in general. Interestingly, this bound is found to be saturated for Landau level systems and fairly tight for the zero-field Chern insulator in twisted semiconductor bilayers near the magic angle~\cite{devakul_magic_2021}. 

In this work, we uncover a universal bound on the structure factor that applies to general chiral topological phases with a quantized Hall response. We show that the static structure factor --- the \textit{equal-time} density-density correlation function --- at long wavelengths has a lower bound solely determined by the many-body Chern number, which governs the Hall response. We discuss applications of our theory to various chiral topological phases, including integer and fractional Chern insulators, topological superconductors, and quantum spin liquids. Our topological bound on structure factor is shown to be useful for Chern insulators in twisted semiconductor bilayers. 
It also provides valuable guidance for identifying topological superconductors and quantum spin liquids from the spin structure factor.

Our finding is derived by considering the response of the system to an external $U(1)$ gauge field, and relies only on the fundamental principle that the energy absorption is always non-negative; this leads to a nontrivial relation between the quantized Hall response and the longitudinal responses. Further relating the longitudinal response to the structure factor by the dissipation-fluctuation theorem, we find a topological bound on the structure factor.

\textit{General theory ---} Consider a two-dimensional (2D) system with a global $U(1)$ symmetry, with corresponding $U(1)$ charge and current densities denoted as $\rho, \vec{j}$. For example, $\rho$ and $\vec{j}$ can represent the electric charge and current, or spin density and spin current in systems with spin $U(1)$ symmetry.
The static structure factor, i.e., the \textit{equal-time} density-density correlation function, is defined as:
\begin{align}
    S_{\vec{q}}&=\frac{1}{V}\expval{\rho_{\vec{q}}{\rho}_{-\vec{q}}} \label{eq:def_Sq}
\end{align} 
with $\rho_{\vec{q}}=\int\dd{\vec{r}}e^{-i\vec{q}\vdot{\vec{r}}}\rho(\vec{r})$ the Fourier transform of the charge density and $V$ the volume. In our convention, $S_{\vec{q}=0}=Vn^2$ with total charge density $n$. By the dissipation-fluctuation theorem, the static structure factor $S_{\vec{q}}$ is related to the density response function $\Pi(\vec{q},\omega)$, which describes the density response to a scalar potential $\phi$, $\delta\rho(\vec{q}, \omega) = \Pi(\vec{q},\omega) \phi(\vec{q},\omega)$:  
\begin{align}
    &-\int_0^{\infty}\dd\omega \Im \Pi(\vec{q}, \omega) = \frac{\pi}{\hbar} S_{\vec{q}}. \label{eq:Pi-S_integrated}
\end{align}

By the continuity equation $\partial \rho/\partial t + \nabla \cdot j =0$, the density response function $\Pi$ can be further related to the current response to the (longitudinal) ``electric'' field associated with the potential $\phi$, 
$\vec{\mathcal{E}}=-\nabla\phi$. %
Denoting the conductivity as $\sigma$, $\Pi$ can be written as  
\begin{align}
    &\Pi(\vec{q},\omega) = -i\frac{q_{\alpha}q_{\beta}\sigma_{\alpha\beta}(\vec{q},\omega)}{\omega} \label{eq:Pi_sigma}. 
\end{align}
Thus combining Eq.~\eqref{eq:Pi-S_integrated} and \eqref{eq:Pi_sigma}, %
we find a relation:
\begin{align}
  q_\alpha q_\beta  \int_0^{\infty} \dd{\omega} \frac{\Re \sigma_{\alpha \beta}(\vec{q}, \omega)}{\omega} = \frac{\pi}{\hbar} S_{\vec{q}}. \label{eq:sigmaS}
\end{align} 

Eq.~\eqref{eq:sigmaS} is a sum rule that relates the negative-first moment of optical conductivity and static structure factor. 
Note that the left hand side of Eq.~\eqref{eq:sigmaS} only involves the real and symmetric part of $\sigma_{\alpha \beta}$, which describes the absorption of linearly polarized light.
In this work, we consider 2D systems with an energy gap, so that optical absorption vanishes at low frequency. This allows us to take $\vec{q}\rightarrow 0$ limit for $\sigma(\vec{q}, \omega)$ and obtain the small-$q$ expansion for $S_{\vec q}$: 
\begin{eqnarray}
S_{\vec{q}} &=& \frac{K_{\alpha\beta}}{2\pi} q_{\alpha}q_{\beta} + ...,  \\
\text{with} \quad K_{\alpha \beta} &=&  2\hbar  \int_0^{\infty} \dd{\omega} \frac{\Re \sigma_{\alpha \beta}(\omega)}{\omega}. 
 \label{eq:W1=K}
\end{eqnarray}
Eq.~\eqref{eq:W1=K} shows that the static structure factor of gapped phases vanishes quadratically at small $\vec q$. The quadratic coefficient $K$ (recently termed quantum weight) characterizes long-wavelength density fluctuation in the ground state. Eq.~\eqref{eq:W1=K} directly relates the quantum weight of a 2D gapped system to the negative-first moment of its optical conductivity~\cite{onishi_quantum_2024-1}.

We now show that the quantum weight has a universal lower bound determined by the quantized dc Hall conductivity or the many-body Chern number. \red{For convenience of presentation, we shall first consider two-dimensional systems with isotropic conductivity tensor $\sigma_{\alpha\beta}=\sigma_{xx}\delta_{\alpha\beta}+\epsilon_{\alpha\beta}\sigma_{xy}$. General cases will be discussed later.} To derive this bound, consider the absorbed power of a 2D system from a monochromatic external field $\vec{\mathcal{E}}(t)=\vec{\mathcal{E}}_0(\omega)e^{-i\omega t} + \mathrm{c.c.}$, which is given by
\begin{align}
    P &= \frac{1}{T}\int \dd{t} \vec{j} \cdot \vec{\mathcal{E}} %
    = \vec{\mathcal{E}}_0(\omega)^*\vdot\vec{j}(\omega) + \mathrm{c.c.} \nonumber \\
    &= \vec{\mathcal{E}}_0^{\dagger} (\sigma(\omega)+\sigma^{\dagger}(\omega)) \vec{\mathcal{E}}_0,
\end{align}
with $T=2\pi/\omega$ the period of $\mathcal{E}(t)$. 
Importantly, the absorbed power $P$ is non-negative for arbitrary $\omega$ and $\vec{\mathcal{E}}_0$. 
In particular, for circularly polarized ``light'' with $x$ and $y$ components equal in magnitude and $90^\circ$ out of phase, $\vec{\mathcal{E}}_0 = (1, \pm i)$, the non-negativity of absorbed power $P\ge 0$ leads to the following inequality~\cite{onishi_fundamental_2024}:
\begin{align}   
    \Re \sigma_{xx}(\omega) \ge \abs{\Im\sigma_{xy}(\omega)}, \label{eq:positivity}
\end{align}
which holds at an arbitrary frequency $\omega$. %

From Eq.~\eqref{eq:positivity} we immediately obtain inequalities between all moments of $\Re \sigma_{xx}(\omega)$ and $\Im \sigma_{xy}(\omega)$: 
\begin{eqnarray}
 \int_0^{\infty} \dd{\omega} \frac{\Re \sigma_{xx}(\omega)}{\omega^n} \geq 
\abs{ \int_0^{\infty} \dd{\omega} \frac{\Im \sigma_{xy}(\omega)}{\omega^n} }.  \label{eq:moments}
\end{eqnarray}
For gapped systems, since there is no absorption at frequency below the gap, the frequency integral in Eq.~\eqref{eq:moments} does not have infrared divergence for any $n$. 
Consider the negative-first moments ($n=1$) of $\Re \sigma_{xx}$ and $\Im \sigma_{xy}$: the former is directly related to the quantum weight by Eq.~\eqref{eq:W1=K}, while the latter is related to the dc Hall conductivity by the Kramers-Kronig relation due to the causality of the response:
\begin{eqnarray}
 \int_0^{\infty} \dd{\omega} \frac{\Im \sigma_{xy}(\omega)}{\omega} = \frac{\pi}{2}\sigma_{xy} (0)
 =\frac{1}{4 \hbar} C. \label{eq:Hall}
\end{eqnarray}
Here, we have used the fact that the dc Hall conductivity of a 2D gapped system is a topological invariant---the many-body Chern number $C$, which may take integer or fractionally quantized values \cite{Thouless1982, niu_quantized_1985}:  
$\sigma_{xy} = C/h$.

Combining Eq.~\eqref{eq:moments} with Eqs.~\eqref{eq:Hall} and \eqref{eq:W1=K}, we obtain a topological bound $K\equiv K_{xx}+K_{yy}\ge \abs{C}$,
or equivalently,
\begin{eqnarray}
S_{\vec q} \ge \frac{\hbar q^2}{2} \abs{\sigma_{xy}}=\frac{q^2}{4\pi}\abs{C}\quad (\vec{q}\to 0). \label{eq:Sq_bound}
\end{eqnarray}
Eq.~\eqref{eq:Sq_bound} applies to general gapped \red{isotropic} 2D systems with $U(1)$ symmetry, including strongly interacting systems such as fractional Chern insulators. The same result holds for lattice models where $\rho_{\vec{q}}=\sum_i e^{i\vec{q}\vdot\vec{r}_i}\rho_i$ is defined with the density operator  $\rho_i$ on site $i$. 
\red{By the same argument, the following bound on the quadratic coefficient of  $S_{\vec{q}}$ holds for general 2D gapped systems (see Supplemental Materials~\footnote{See Supplemental Material for more detailed calculations, which includes Refs.~\cite{volovik_fractional_1989, senthil_spin_1999} \label{SM}}):
\begin{align}
    K_{xx}+K_{yy} \ge \abs{C}.
\end{align}
Note that in the thermodynamic limit $V\to\infty$ considered throughout this work, $S_{\vec{q}}$ defined by Eq.~\eqref{eq:def_Sq} is proportional to the volume $V$ at reciprocal lattice vectors (including $\vec{q}=0$), whereas it is an intensive quantity of $\order{1}$ at generic $\vec{q}$. Thus, $S_{\vec{q}}$ in $\vec{q}\to 0$ limit differs from $S_{\vec{q}=0}=Vn^2$. 
}  Below, we will apply Eq.~\eqref{eq:Sq_bound} to various systems: Chern insulators, topological superconductors and chiral spin liquids.

{\it Chern insulators --- }  
Let us first consider Chern insulators, where the $U(1)$ symmetry corresponds to  
electric charge conservation. In this case, $\rho$ and $\vec{j}$ correspond to the electron number density and current respectively.  
The electric conductivity $\sigma^e$ is related to the conductivity $\sigma$ in our general theory as $\sigma^e=e^2\sigma$, and the many-body Chern number $C$ governs the quantized electrical Hall conductivity $\sigma^e_{xy}=C e^2/h$. Eq.~\eqref{eq:Sq_bound} imposes a lower bound on the charge structure factor for Chern insulators: 
$S_{\vec q} \ge q^2\abs{C}/(4\pi)$ at small $q$.  

For noninteracting insulators, the inequality~\eqref{eq:Sq_bound} can be explicitly verified. The structure factor is given by
\begin{align}
    S_{\vec{q}} &= \int_{\rm BZ}\frac{\dd^d{\vec{k}}}{(2\pi)^d} \Tr[P(\vec{k})(P(\vec{k})-P(\vec{k}+\vec{q}))] \label{eq:BI_Sq} \\
    &= \frac{q_\alpha q_\beta}{2\pi}\int_{\rm BZ}\frac{\dd^2{\vec{k}}}{2\pi} g_{\alpha\beta}(\vec{k}) \quad (\vec{q}\to 0),  \label{K-BI}
\end{align}
where $P(\vec{k})=\sum_{n}^{\rm occ} \ket{u_{n\vec{k}}}\bra{u_{n\vec{k}}}$ is the projection operator onto the occupied bands at wavevector $\vec{k}$ with $\ket{u_{n\vec{k}}}$ the cell-periodic Bloch wavefunction for $n$-th band. \red{See Supplemental Materials~\cite{Note1} for the derivation of Eq.~\eqref{eq:BI_Sq}.} 
$g_{\mu\nu}(\vec{k})=\Re Q_{\mu\nu}(\vec{k})$ is the quantum metric with $Q_{\mu\nu}(\vec{k})\equiv\sum_{n}^{\rm occ}\mel{\partial_{\mu}u_{n\vec{k}}}{(1-P(\vec{k}))}{\partial_{\nu}u_{n\vec{k}}}$ the quantum geometric tensor of the occupied Bloch bands.
Eq.~\eqref{K-BI} shows that the quadratic coefficient of structure factor of band insulators, $K=K_{xx}+K_{yy}$, reduces to the integral of the quantum metric over the Brillouin zone, which is related to the spread of Wannier function~\cite{marzari_maximally_1997} and the optical response~\cite{souza_polarization_2000, komissarov_quantum_2024, onishi_quantum_2024-1}. Because $g_{xx}+g_{yy}\ge \abs{\Omega_{xy}}$ with the Berry curvature $\Omega_{\mu\nu}=-2\Im Q_{\mu\nu}$~\cite{roy_band_2014},  $K\ge \abs{C}$ follows \cite{peotta_superfluidity_2015, mera_kahler_2021, ledwith_vortexability_2022}, consistent with our general result~\eqref{eq:Sq_bound} in noninteracting cases.
This inequality is related to the fact that the Wannier functions in Chern bands cannot be exponentially localized~\cite{brouder_exponential_2007} and must have a minimum spread~\cite{marzari_maximally_1997, peotta_superfluidity_2015}. 

To illustrate our topological bound on structure factor, we consider a two-band $k\cdot p$ model described by the following Hamiltonian:
\begin{align}
    &H = \sum_{\vec{k}} \begin{pmatrix}
        \cre{c}_{\vec{k}\up} & \cre{c}_{\vec{k}\down}
    \end{pmatrix} H(\vec{k})\begin{pmatrix}
        \ani{c}_{\vec{k}\up} \\
        \ani{c}_{\vec{k}\down}
    \end{pmatrix}, \label{eq:Chern_Ham} \\
    &H(\vec{k}) = \begin{pmatrix}
        \xi(\vec{k}) & \Delta(\vec{k}) \\
        \Delta(\vec{k})^* & -\xi(\vec{k})
    \end{pmatrix},
\end{align}
where $\ani{c}_{\vec{k}\up(\down)}$ is the annihilation operator of an electron with wavevector $\vec{k}$ and spin up (down) and $\xi(\vec{k}) = k^2/(2m) - \mu$. The hybridization term is either $s$-wave, $\Delta(\vec{k})=\Delta_s$, or $p$-wave, $\Delta(\vec{k})=\Delta_p(k_x-ik_y)$. For the $p$-wave case, the system becomes a Chern insulator with $\abs{C}=1$ when $\mu>0$ and is trivial for $\mu<0$. At $\mu=0$, the system becomes gapless and undergoes the topological phase transition. 

\begin{figure}
    \centering    
    \includegraphics[width=\columnwidth]{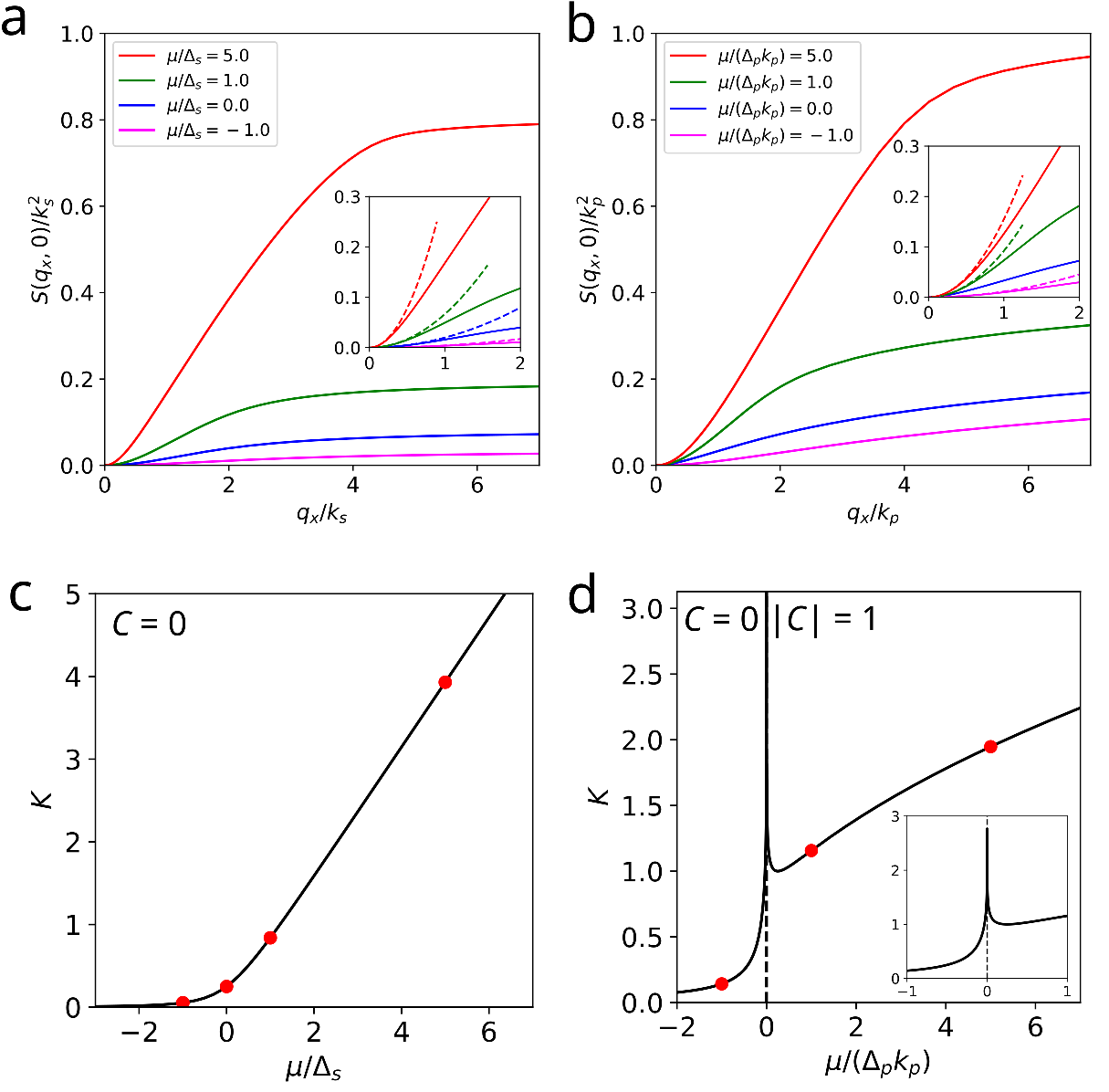}
    \caption{(a, b) The structure factor $S(q_x, 0)$ for the Hamiltonian~\eqref{eq:Chern_Ham} when $\Delta(\vec{k})$ is (a) $s$-wave and (b) $p$-wave. The dashed lines are the quadratic curves determined by the quantum weight. (c, d) The quantum weight $K$ as a function of $\mu$ for (c) $s$-wave and (d) $p$-wave. Here, $k_s=(2m\Delta_s)^{1/2}$ and $k_p=2m\Delta_p$.
    The results are identical to the spin structure factor for the superconductors described by~\eqref{eq:SC_Ham}.}
    \label{fig:topo_SC}
\end{figure}

The structure factor $S_{\vec q}$ and the quantum weight $K$ for both $s$-wave and $p$-wave $\Delta(\vec{k})$ for several $\mu$ are shown in Fig.~\ref{fig:topo_SC}. In all the cases except at the gapless point $\mu=0$ in the $p$-wave case, the structure factor is quadratic in $q$ near $q=0$, with the coefficient given by the quantum weight $K$ (dashed line in the insets). In $s$-wave cases, $K$ continuously decreases as $\mu$ is reduced, and vanishes as $\mu \rightarrow -\infty$. 

In contrast, for $p$-wave case, the quantum weight $K$ in the topological phase ($\mu>0$) is always larger than or equal to $\abs{C}=1$, being consistent with our general result~\eqref{eq:Sq_bound}. Interestingly, when $\mu = m\Delta_p^2/2$, the energy dispersion is parabolic as $\ene_{\pm}(\vec{k}) = \pm\qty(k^2/(2m) + \mu)$ and the quantum weight $K$ is equal to $1$ ~\cite{hu_fractional_2018, tam_kane_2024}, saturating its topological bound. 
At the topological phase transition $\mu=0$, $K$ has a logarithmic divergence $\sim\log\abs{\mu}$ as previously shown~\cite{onishi_fundamental_2024}.  
See Supplemental Materials~\cite{Note1} for more details.

The topological bound on charge structure factor is also saturated for a particular type of Chern insulators, quantum Hall states in Landau level systems. For 2D electron systems with parabolic dispersion $p^2/(2m)$ under a magnetic field $B$, the static structure factor at long wavelength is given by
\begin{align}
S^{\rm LL}_{\vec{q}} = \nu q^2/(4\pi) + \dots. \label{eq:SLL}
\end{align}
where $\nu=n/(2\pi l^2)$ is the Landau level filling factor with $n$ the carrier density and $l=\sqrt{\hbar/(eB)}$ is the magnetic length. 
This follows from the fact that the optical spectral weight is exhausted by Kohn's mode at cyclotron frequency $\omega_c=eB/m$~\cite{kohn_cyclotron_1961, girvin_magneto-roton_1986, girvin_modern_2019} and the relation between structure factor and optical conductivity~\eqref{eq:W1=K}. %
The Galilean invariance also dictates the quantized Hall conductivity $|C|=\nu$. 
Importantly, Eq.~\eqref{eq:SLL} only relies on Kohn's theorem and therefore holds for both integer and fractional quantum Hall states.  Therefore, quantum Hall states in Landau level systems saturate our general bound on the structure factor.
Note that the structure factor $S^{\rm LL}_{\bm q}$
is different from the guiding center structure factor $\bar{S}_{\bm q}$ or the Landau-level-projected density-density correlation function; the latter is proportional to $q^4$~\cite{girvin_magneto-roton_1986}.

\begin{figure}[tbp]
    \centering
    \includegraphics[width=1.0\columnwidth]{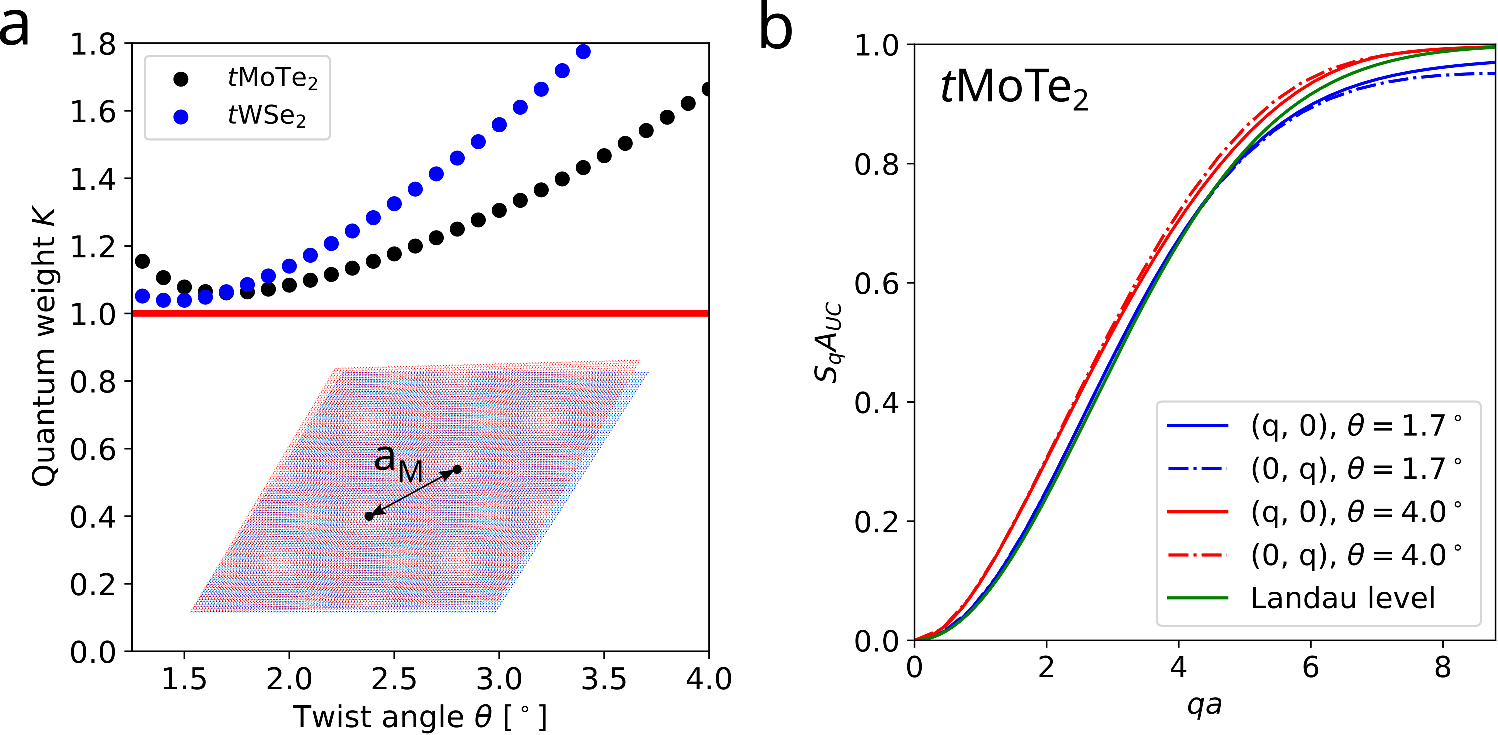}
    \caption{(a) Quantum weight for $t$MoTe$_2$ and $t$WSe$_2$ at filling factor $\nu=-1$. (b) The static structure factor $S_q$ for $t$MoTe$_2$ at $\theta=\ang{1.7}, \ang{4.0}$ and the lowest Landau level. $S_q$ is scaled with the unit cell area $A_{UC}=\sqrt{3}a^2/2$. For $t$MoTe$_2$, $a$ is moir\'e lattice constant $a_M$. For Landau level, $a$ is defined so that $A_{UC} = \sqrt{3} a^2/2$. See Supplemental Materials~\cite{Note1} for details.}
    \label{fig:MoTe2_WSe2}
\end{figure}

Chern insulators in twisted transition metal dichalcogenides provide an excellent platform to test our theory. Following theoretical proposals~\cite{devakul_magic_2021, li_spontaneous_2021, crepel_anomalous_2023}, integer and fractional quantum anomalous Hall effect have been experimentally observed in twisted MoTe$_2$ ($t$MoTe$_2$) and WSe$_2$ ($t$WSe$_2$)~\cite{park_observation_2023, cai_signatures_2023, xu_observation_2023, zeng_thermodynamic_2023, foutty_mapping_2023}. Our general bound~\eqref{eq:Sq_bound} applies to both integer and fractional Chern insulators in these materials. Here, we explicitly calculate quantum weight and structure factor of the integer Chern insulator at $\nu=1$ where charge carriers completely fill the lowest Chern band of one spin~\cite{wu_topological_2019, devakul_magic_2021, reddy_fractional_2023} (assuming that band mixing is negligible). The results show the inequality~\eqref{eq:Sq_bound} always holds and is fairly tight: $K=1.07$ at $\theta=\ang{1.7}$ for $t$MoTe$_2$, and $K=1.04$ at $\theta=\ang{1.4}$ for $t$WSe$_2$, being close to the bound $\abs{C}=1$. Interestingly, the structure factor of $t$MoTe$_2$ is nearly isotropic and close to that of the lowest Landau level, especially at $\theta=\ang{1.7}$, indicating that the system is close to the lowest Landau level.  

{\it Quantum spin Hall insulators} --- The topological bound for Chern insulators can be straightforwardly extended to quantum spin Hall insulators. In time-reversal invariant insulators with $s^z$ conservation, 
\red{we can define the spin Chern number by introducing the electric field $\vec{E}_{\uparrow}$ that is coupled only to spin-$\uparrow$ particles. The Hall current $\vec{j}$ induced by $\vec{E}_{\uparrow}$ defines the corresponding conductivity and Chern number: $j_y=\sigma_{yx}E_{\uparrow,x}$, $\sigma_{yx}=C_{\uparrow}e^2/h$. Similarly, the Chern number for spin-$\downarrow$ particles $C_{\downarrow}$ is defined, and the time-reversal symmetry dictates $C_{\uparrow}=-C_{\downarrow}\equiv C_{s}$ with $C_s$ the spin Chern number. For each spin sector, the structure factor $S^{\sigma}_{\vec{q}}$ ($\sigma=\uparrow,\downarrow$) satisfies the topological bound~\eqref{eq:Sq_bound}. In noninteracting cases, the charge static structure factor is given by the sum of the ones for each spin sector and thus is bounded by $2\abs{C_s}$: $S_{\vec{q}} \ge q^2\abs{C_s}/(2\pi)$.}

{\it Spin $U(1)$ symmetry} ---%
We can also apply our general theory to systems with spin $U(1)$ symmetry.  
When the total spin along $z$ direction $s^z$ is conserved, we can relate $\rho$ and $\vec{j}$ to the spin density $s^z(\vec{r})$ and the spin current $\vec{j}^s(\vec{r})$ as 
$s^z(\vec{r}) \equiv 
s_0\rho(\vec{r})$
and
$\vec{j}^s(\vec{r}) \equiv 
s_0\vec{j}(\vec{r})$. Here, $s_0$ has a dimension of spin angular momentum. %
The potential $\phi$ in this case corresponds to the Zeeman field, $-s_0B_z(\vec{r})$, which couples to the spin density as $\rho\phi = -B_z s^z$. The corresponding ``electric'' field is the gradient of the magnetic field, $\vec{\mathcal{E}}= s_0\nabla B_z$. By defining the spin conductivity as $\vec{j}^s = \sigma^s \nabla B_z$, the conductivity in our general theory $\sigma$ is related to $\sigma^s$ as $\sigma^s = s_0^2 \sigma$, while the static structure factor is related to the spin static structure factor $S^s_{\vec{q}} \equiv \expval{\hat{s}^z_{\vec{q}}\hat{s}^z_{-\vec{q}}} /V$ as $S^s_{\vec{q}} = s_0^2 S_{\vec{q}}$. 

For systems with a spin gap, %
we can %
apply our general theory %
to obtain a bound on the spin static structure factor in terms of the Hall conductivity for spin:
\begin{align}
    S^s_{\vec{q}} &\ge \frac{\hbar q^2}{2}\abs{\sigma^s_{xy}}
    =\frac{q^2s_0^2}{4\pi}\abs{C} 
    \quad (\vec{q}\to 0). \label{eq:Ss_sigmas}
\end{align}
In the following, we examine two examples with the spin $U(1)$ symmetry 
: chiral topological superconductors and chiral spin liquids.

{\it Chiral topological superconductors ---}
For superconductors %
with a full superconducting gap and spin $U(1)$ symmetry, the spin conductivity vanishes at frequencies below the gap. This allows us to apply our theory to the spin structure factor.

As an example, we consider a superconductor with $s_z$ conservation described by the following mean-field Hamiltonian:
\begin{align}
    H &= \sum_{\vec{k}} \begin{pmatrix}
        \cre{c}_{\vec{k}\up} & \ani{c}_{-\vec{k}\down}
    \end{pmatrix} H(\vec{k}) \begin{pmatrix}
        \ani{c}_{\vec{k}\up} \\
        \cre{c}_{-\vec{k}\down}
    \end{pmatrix} 
    \equiv \sum_{\vec{k}} 
    \cre{\Psi}_{\vec{k}} H(\vec{k})    %
    \ani{\Psi}_{\vec{k}}.
    \label{eq:SC_Ham}
\end{align}
where $H(\vec k)$ takes the same form as the two-band model ~\eqref{eq:Chern_Ham} for insulators.  
As a mean-field Hamiltonian for superconductors,  $\Delta(\vec{k})$ represents the pairing function; $\Delta(\vec{k})=\Delta_s$ describes a $s$-wave pairing while $\Delta(\vec{k})=\Delta_p(k_x-ik_y)$ describes a $p$-wave pairing \cite{read_paired_2000}. The $p$-wave pairing considered here is a spin triplet with zero spin angular momentum in $z$-direction. The parameter $\mu$ describes the evolution of the superconducting state from BCS ($\mu>0$) to BEC ($\mu<0$) regime. 

Since the Hamiltonian conserves $U(1)$ charge $\cre{\Psi}\ani{\Psi}$, we can apply our theory to the structure factor for the density $\cre{\Psi}\ani{\Psi}$. Importantly, $\cre{\Psi}\ani{\Psi}$ corresponds to the spin density $s^z$%
, the structure factor corresponds to the spin structure factor, and Eq.~\eqref{eq:Ss_sigmas} holds with $s_0=\hbar/2$. Therefore, the spin structure factor of $s_z$-conserving superconductors is (up to a factor of $s_0^2$) identical to the charge structure factor for insulators, which is shown in Fig.~\ref{fig:topo_SC}.  For the chiral $p$-wave topological superconductor, the spin structure factor and the corresponding quantum weight have a lower bound by the Chern number. As another example, we discuss $d+id$-wave spin-singlet superconductors on a lattice in Supplemental Materials~\cite{Note1}.

{\it Chiral spin liquids} --- Our theory also applies to chiral spin liquids with spin $U(1)$ symmetry ~\cite{kalmeyer_theory_1989}. For example, consider a spin-$1/2$ Heisenberg model on a lattice with total spin $s^z$ conserved. 
The $U(1)$ symmetry becomes more clear when the model is mapped to a tight-binding model of hard-core bosons: a spin-up state at site $i$ is represented by one boson at site $i$, with an infinite onsite repulsion preventing more than one bosons on the same  site.
Formally, this mapping is achieved by rewriting the spin operator $\vec{s}_i=(\hbar/2)\vec{\sigma}$ with $\ani{a}_{i} \equiv \qty(s^x_i -is^y_{i})/\hbar$. The total number of bosons is conserved due to the spin $U(1)$ symmetry, i.e., total $s^z$ conservation.

The $U(1)$ gauge field couples to the density $\rho_i = \cre{a}_i\ani{a}_i$
and the corresponding current $\vec{j}$, determined by the derivative of the Hamiltonian with respect to the gauge field. 
Since one boson carries spin of $\hbar$, the natural choice for the coupling constant is $s_0=\hbar$, and the spin density and the spin current are related to the density and the current of the bosons as $s^z_i=\hbar(\rho_i-1/2), \vec{j}^s = \hbar\vec{j}$.

With these, we can apply the topological bound~\eqref{eq:Ss_sigmas}. The quantum weight $K$ and the Chern number $C$ are related to the spin structure factor and the spin conductivity as $S^s_{\vec{q}} = \hbar^2q^2K/(4\pi)$, $\sigma^s_{xy} = \hbar C/(2\pi)$, and the topological bound~\eqref{eq:Ss_sigmas} reduces to the following relation between $S^s_{\vec{q}}$ and the Chern number $C$:
\begin{align}
    S^s_{\vec{q}} &\ge \frac{\hbar^2}{4\pi}q^2 \abs{C} \quad (\vec{q}\to 0) \label{eq:QSL_bound}
\end{align}
This analysis is readily generalized to spin-$S$ systems. 

\textit{Discussion} --- While our work focuses on the topological bound on the structure factor which is a ground state property, it is also closely related to the topological bound on the energy gap shown recently~\cite{onishi_fundamental_2024}. In fact, the upper bound on the energy gap $\Delta$ can be obtained from our lower bound on the structure factor by using
the Feynmann-Bijl formula~\cite{feynman_atomic_1954}:
\begin{align}
    \Delta \le \frac{\expval{\comm{\rho_{\vec{q}}}{\comm{H}{\rho_{-\vec{q}}}}}}{2VS_{\vec{q}}}, \label{eq:FB-formula}
\end{align}
where $\expval{\dots}$ denotes the expectation value in the ground state.
The derivation of Eq.~\eqref{eq:FB-formula} for general $U(1)$ symmetric systems is given in Supplemental Materials~\cite{Note1}. Using the continuity equation $\dot{\rho}_{\vec{q}} + i\vec{q}\vdot\vec{j}_{\vec{q}}=0$ and the relation $\vec{j}_{\vec{q}}=-\pdv*{H}{\vec{A}_{-\vec{q}}}$, we can rewrite Eq.~\eqref{eq:FB-formula} as 
\begin{align}
    \Delta \le \frac{\hbar^2 q_{\mu}q_{\nu}}{2VS_{\vec{q}}}\expval{\pdv[2]{H}{A^{\mu}_{\vec{q}}}{A^{\nu}_{-\vec{q}}}}, \label{eq:FB-formula_2} 
\end{align}
By taking $\vec{q}\to0$ limit and applying the topological bound on the structure factor~\eqref{eq:Sq_bound}, we find 
\begin{align}
    \Delta &\le \frac{\pi \hbar^2}{K}\expval{\frac{1}{V}\pdv[2]{H}{\vec{A}}} 
    \le \frac{\pi \hbar^2}{\abs{C}}\expval{\frac{1}{V}\pdv[2]{H}{\vec{A}}}, \label{eq:gap_bound}
\end{align}
where $\vec{A}$ is the uniform $U(1)$ vector potential.
Eq.~\eqref{eq:gap_bound} is the topological gap bound that applies to general topological phases with a finite Chern number. In particular, when the Hamiltonian is quadratic in the momentum $\vec{p}$ of the particles, $H=\sum_i (\vec{p}_i-\vec{A})^2/(2m) + \dots$, 
Eq.~\eqref{eq:gap_bound} reduces to the previously derived topological gap bound for Chern insulators: $\Delta\le 2\pi\hbar^2n/(m\abs{C})$ with the charge density $n$. It is worth noting that Eq.~\eqref{eq:gap_bound} also applies to systems with spin-$U(1)$ symmetry as we discussed in this work. For $s_z$-conserving superconductors, the gap $\Delta$ is the spin gap, i.e., the energy of spin excitation at long wavelength $\vec q\rightarrow 0$. The upper bound on the spin gap of topological superconductors also gives an upper bound on the superconducting gap.

Since the structure factor represents density fluctuations, the existence of a lower bound shows that nontrivial topology necessarily requires quantum fluctuation in the $U(1)$ charge density. 
The static structure factor is a physical observable for all systems with $U(1)$ symmetry and can be measured by X-ray or neutron scattering, or determined by the optical conductivity measurement through the sum rule~\eqref{eq:sigmaS}. Since our bound applies to interacting systems, it will be interesting to investigate fractional Chern insulators in twisted semiconductor bilayers and multilayer graphene/h-BN numerically and experimentally.

Our topological bound on the structure factor may be useful in identifying topological superconductors and quantum spin liquids in experiments. If the spin structure factor is smaller than the lower bound~\eqref{eq:QSL_bound} determined by a Chern number $C$, we can rule out the possibility of a chiral spin liquid with the Chern number $C$. A large spin structure factor at small $\vec q$ is a necessary criterion for a system to be considered as a \red{$U(1)$ chiral} quantum spin liquid candidate.

While this work focuses on the bound in topological phases characterized by the Chern number, an interesting research direction is to find the topological bound in other topological phases of matter. \red{We also note that the generalization to three-dimensional systems with Coulomb interaction is nontrivial because the density response $\Pi$ and the optical conductivity $\sigma$ at $\vec{q}\to 0$ are not related as Eq.~\eqref{eq:Pi_sigma}, as discussed in details in Ref.~\cite{onishi_quantum_2024-1}.} 

In conclusion, we have proved a topological bound on static structure factor in general systems with $U(1)$ symmetry. Our derivation relies only on basic physical principles and \red{the bound applies to a wide range of two-dimensional gapped} systems, such as integer or fractional Chern insulators, topological superconductors, and quantum spin liquid, revealing universal information about correlation function obtained from topology.

\begin{acknowledgements}
This work was supported by the U.S. Army Research Laboratory and the U.S. Army Research Office through the Institute for Soldier Nanotechnologies under Collaborative Agreement Number W911NF-18-2-0048 and a Simons Investigator Award from the Simons Foundation. YO is grateful for the support provided by the Funai Overseas Scholarship. LF was partly supported by the Air Force Office of Scientific Research under award number FA2386-24-1-4043. 
\end{acknowledgements}

\bibliography{references, private_comm}

\appendix
\begin{widetext}
\section{Supplemental Materials}
\section{Generalization to anisotropic systems}
The topological bound on the structure factor can be generalized to general anisotropic systems. 
In the case of anisotropic systems, the conductivity tensor $\sigma_{\alpha\beta}(\omega)$ is not of the form $\sigma_{\alpha\beta}=\sigma_{xx}\delta_{\alpha\beta} + i\epsilon_{\alpha\beta}\sigma_{xy}$. Therefore, the inequality (8) in the main text due to the non-negativity of absorbed power $P\ge 0$ is generalized to the following inequality:
\begin{align}
    \Re \sigma_{xx}(\omega) + \Re\sigma_{yy}(\omega) \ge 2\abs{\Im\sigma_{xy}(\omega)}.
\end{align}
With this change, we can derive the topological bound on the quantum weight $K_{\alpha\beta}$ and the structure factor as 
\begin{align}
    &K_{xx} + K_{yy} \ge \abs{C}, \\
    &S_{(q,0)} + S_{(0, q)} \ge \hbar q^2\abs{\sigma_{xy}} = \frac{q^2}{4\pi}\abs{C} \quad (\vec{q}\to 0).
\end{align}

\section{Calculation details for model of Chern insulators and topological superconductors}
Here we present details of the calculation for the model discussed in the main text:
\begin{align}
    H &= \sum_{\vec{k}} \begin{pmatrix}
        \cre{c}_{\vec{k}\up} & \cre{c}_{\vec{k}\down}
    \end{pmatrix} H(\vec{k})\begin{pmatrix}
        \ani{c}_{\vec{k}\up} \\
        \ani{c}_{\vec{k}\down}
    \end{pmatrix}, \label{eq:Chern_Ham_ap} \\
    H(\vec{k}) &= \begin{pmatrix}
        \xi(\vec{k}) & \Delta(\vec{k}) \\
        \Delta(\vec{k})^* & -\xi(\vec{k})
    \end{pmatrix},
\end{align}
where $\xi(\vec{k}) = \hbar^2k^2/(2m) - \mu$.
If the annihilation and creation operators are appropriately replaced, this Hamiltonian will be a model for superconductors as well. When $\Delta(\vec{k}) = \Delta_s$ (constant), the system is topologically trivial, while the system can be topological with Chern number $\abs{C} = 1$ for $\Delta(\vec{k}) = \Delta_p (k_x-ik_y)$. We shall refer to the former case as $s$-wave and the latter as $p$-wave. 

The quantum weight of this system can be calculated with the quantum geometric tensor. Here we first present the general expression for the quantum geometric tensor for the two-band systems, and then we apply it to the specific case of the $s$-wave and $p$-wave.

\subsection{General expression of the quantum geometric tensor for two-band systems}
Consider general Hamiltonian given by 
\begin{align}
    H(\vec{k}) = \vec{d}(\vec{k})\vdot \vec{\sigma},
\end{align}
with $\vec{\sigma}$ is the vector of Pauli matrices. 
The energy dispersion is given by 
\begin{align}
    \ene_{\pm}(\vec{k}) &= \pm \abs{\vec{d}(\vec{k})}
\end{align}
The Bloch wavefunction $\ket{u_{\vec{k},\pm}}$ is determined by a unit vector $\vec{n}(\vec{k})=\vec{d}(\vec{k})/\abs{\vec{d}}(\vec{k})=(\sin\theta\cos\phi, \sin\theta\sin\phi, \cos\theta)$ with $\theta$ and $\phi$ functions of $\vec{k}$, and is given by 
\begin{align}
    &\ket{u_{\vec{k},+}} = \begin{pmatrix}
        \cos(\theta/2) \\
        e^{i\phi}\sin(\theta/2)
    \end{pmatrix} \\
    &\ket{u_{\vec{k},-}} = \begin{pmatrix}
        \sin(\theta/2) \\
        -e^{i\phi}\cos(\theta/2)
    \end{pmatrix}
\end{align}
The quantum geometric tensor $Q_{\mu\nu}^{\pm} = g_{\mu\nu} \mp (i/2)\Omega_{\mu\nu}$ for each band $(\pm)$ is then given by 
\begin{align}
    g_{\mu\nu} &= \frac{1}{4}(\partial_\mu\vec{n})\vdot(\partial_\nu\vec{n}) = \pdv{\theta}{k_{\mu}}\pdv{\theta}{k_{\nu}} g_{\theta\theta} + \pdv{\phi}{k_{\mu}}\pdv{\phi}{k_{\nu}} g_{\phi\phi} \nonumber \\
    &= \frac{1}{4}\pdv{\theta}{k_{\mu}}\pdv{\theta}{k_{\nu}} + \frac{\sin^2\theta}{4}\pdv{\phi}{k_{\mu}}\pdv{\phi}{k_{\nu}} \\
    \Omega_{\mu\nu} &= -\frac{1}{2}\vec{n}\vdot(\partial_\mu\vec{n}\cross\partial_\nu\vec{n}) = \qty(\pdv{\theta}{k_{\mu}}\pdv{\phi}{k_{\nu}} - \pdv{\phi}{k_{\mu}}\pdv{\theta}{k_{\nu}}) \Omega_{\theta\phi} \nonumber \\
     &= \frac{\sin\theta}{2}\qty(\pdv{\theta}{k_{\mu}}\pdv{\phi}{k_{\nu}} - \pdv{\phi}{k_{\mu}}\pdv{\theta}{k_{\nu}})
\end{align}
Now let us apply these formulas to the $s$-wave case and $p$-wave case of the Hamiltonian~\eqref{eq:Chern_Ham_ap}.

\subsection{$s$-wave case}
The Hamiltonian for $s$-wave can be written as $H(\vec{k}) = \xi_k \sigma_z + \Delta_s \sigma_x$ with $\sigma_{\mu}$ is the Pauli matrices. Thus the energy dispersion is given by 
\begin{align}
    \ene_{\pm}(\vec{k}) &= \pm \sqrt{\xi_k^2 + \Delta_s^2} \equiv \pm \ene(\vec{k}).
\end{align}
The Bloch wavefunction $\ket{u_{\vec{k},\pm}}$ is given by 
\begin{align}
    \ket{u_{\vec{k},+}} = \begin{pmatrix}
        \cos (\theta_k/2) \\
        \sin (\theta_k/2)
    \end{pmatrix} \\
    \ket{u_{\vec{k}, -}} = \begin{pmatrix}
        \sin (\theta_k/2) \\
        -\cos (\theta_k/2)
    \end{pmatrix}
\end{align}
where $\theta_k$ is determined so that it satisfies 
\begin{align}
    \cos\theta_k = \xi_k/\ene(\vec{k}), \\
    \sin\theta_k = \Delta_s/\ene(\vec{k}). 
\end{align}

The quantum weight $K=K_{xx}+K_{yy}$ can be calculated with the expression with the quantum metric (Eq.~(13) in the main text):
\begin{align}
    K = \int\frac{\dd^2k}{2\pi} \qty(g_{xx} + g_{yy}).
\end{align}
Here, the integrand $g_{xx}+g_{yy}$ is given by 
\begin{align}
    g_{xx}+g_{yy} &= \frac{\Delta_s^2 \hbar^4}{4m^2} \frac{k^2}{\ene_k^4}
\end{align}
and thus the quantum weight is
\begin{align}
    K &= \frac{1}{2}\int_0^{\infty} \dd{x} \frac{x}{\qty((x-(\mu/\Delta_s))^2 + 1)^2} \nonumber \\
     &= \frac{1}{8} \qty(2 + \frac{\mu}{\Delta_s}(\pi + \arctan(\mu/\Delta_s)))
\end{align}

\subsection{$p$-wave case}
The $\vec{d}(\vec{k})$ in this case is given by $\vec{d}(\vec{k}) = (\xi_{\vec{k}}, \Delta_p k_x, \Delta_p k_y)$. The energy dispersion is given by 
\begin{align}
    \ene_{\pm}(\vec{k}) = \pm\sqrt{\xi_{\vec{k}}^2 + \Delta_p^2 k^2}.
\end{align}
The quantum weight $K$ is given by 
\begin{align}
    K = \frac{1}{4}\int_{0}^{\infty} x\dd{x} \frac{2y^2 + 2x^4 + x^2}{\qty{\qty(y- x^2)^2 + x^2}^2}
\end{align}
where $y = \hbar^2 \mu/(2m \Delta_p^2)$. 

Interestingly, when $y=1/4$, i.e., $\mu = m\Delta_p^2/(2\hbar^2)$, the energy dispersion is exactly parabolic:
\begin{align}
    \ene_{\pm}(\vec{k}) = \pm\qty(\frac{\hbar^2k^2}{2m} + \mu),
\end{align}
and $K$ can be analytically calculated and is equal to 1:
\begin{align}
    K(\mu = m\Delta_p^2/(2\hbar^2)) &= 1.
\end{align}
Namely, the quantum weight saturates the bound determined by the Chern number $\abs{C}=1$.

\section{Quantum geometry in $U(1)$ symmetric systems}
We consider ground states with degeneracy $r$ in the $U(1)$ symmetric system. Under twisted boundary condition, the ground states satisfies 
\begin{align}
    &\Psi_{a\vec{\kappa}}(\vec{r}_1, \dots, \vec{r}_i + \vec{L}_{\mu}, \dots, \vec{r}_N) \nonumber \\
    &=  e^{i\vec{\kappa}\vdot\vec{L}_{\mu}}\Psi_{a\vec{\kappa}}(\vec{r}_1, \dots, \vec{r}_i, \dots, \vec{r}_N), \label{eq:TBC}
\end{align}
where $\Psi_{a\vec{\kappa}}$ is the $a$-th ground state wavefunction in the first quantization representation, and $\vec{r}_i$ specifies the position of $i$-th particle corresponding to the conserved $U(1)$ charge. $\vec{L}_{\mu}=(0,\dots, L_{\mu}, \dots,0)$ specifies the system size in $\mu$-direction $L_{\mu}$, and the vector $\vec{\kappa}$ specifies the twisted boundary condition. Note that the number of particle is a good quantum number due to the global $U(1)$ symmetry. 

Introducing the twisted boundary condition $\vec{\kappa}$ is equivalent to coupling the system to a uniform $U(1)$ gauge field, $\vec{A}=\vec{\kappa}$. Therefore, the current response to the gauge field can be described by the quantum geometric quantity associated with the twisted boundary condition $\vec{\kappa}$, and the corresponding conductivity for the $U(1)$ charge with the twisted boundary condition $\vec{\kappa}$ $\sigma(\omega;\kappa)$ is given by
\begin{align}
    &\sigma_{\mu\nu}(\omega; \kappa) = \frac{1}{\hbar} \sum_{n, m} \frac{-iE_{nm} A^{\mu}_{nm}A^{\nu}_{mn} f_{nm}}{\hbar\omega + E_{nm} + i\delta}, \label{eq:interacting_sigma}
\end{align}
where $E_{nm}=E_n-E_m$ is the energy difference between $n$-th and $m$-th many body eigenstate, and $\partial_{\mu}$ is the derivative with respect to $\kappa_{\mu}$. $f_{nm}=f_n-f_m$ with the probability $f_n$ that $n$-th eigenstate is realized. At zero temperature, the canonical distribution gives $f_n = 1/r$ when the state $n$ is one of the $r$-fold degenerated ground states and otherwise $f_n=0$.
$A^{\mu}_{nm}=\mel{n,\vec{\kappa}}{i\partial_{\mu}}{m,\vec{\kappa}}$ is the interband Berry connection for the $n$-th and $m$-th eigenstates the interacting system under the boundary condition $\vec{\kappa}$. 

Further we introduce the quantum geometric tensor $Q^{\mu\nu}$ for the many-body ground states as 
\begin{align}
    Q^{\mu\nu}_{ab} &= \mel{\partial_{\mu}\Psi_{a\vec{\kappa}}}{(1-P_{\vec{\kappa}})}{\partial_{\nu}\Psi_{b\vec{\kappa}}}.
\end{align}
where $\partial_{\mu}$ refers to the derivative with respect to $\kappa_{\mu}$, and $P_{\vec{\kappa}}$ is the projection operator onto the ground state subspace for the boundary condition $\vec{\kappa}$.
$Q^{\mu\nu}$ is in general an $r\times r$ matrix and non-Abelian quantum geometric  tensor. The symmetric and antisymmetric components of $Q^{\mu\nu}$ define the quantum metric and Berry curvature for the many-body states $\ket{\Psi_{\vec{\kappa}}}$ respectively: $Q^{\mu\nu} = G^{\mu\nu} - iF^{\mu\nu}/2$. The trace of them gives the Abelian quantum geometry for the entire ground state subspace: $g^{\mu\nu}_{\rm tot}\equiv\Tr G^{\mu\nu}, \Omega^{\mu\nu}_{\rm tot}\equiv\Tr F^{\mu\nu}$. 
If the ground state is unique, the quantum geometric tensor reduces to the Abelian quantum geometric tensor. 

Since the relation between the conductivity $\sigma(\omega)$ and the Berry connection $A^{\mu}$ is identical to the electric conductivity~\cite{onishi_fundamental_2024}, we can readily show the sum rule for the negative-first moment:
\begin{align}
    \int_0^{\infty}\dd{\omega} \frac{\Re \sigma_{\mu\nu}(\omega)}{\omega} = \frac{1}{2\hbar}g_{\mu\nu} \label{eq:SWM_ap}\\
    \int_0^{\infty}\dd{\omega} \frac{\Im \sigma_{\mu\nu}(\omega)}{\omega} = \frac{1}{4\hbar}\Omega_{\mu\nu} \label{eq:KK_ap}
\end{align}
where $g_{\mu\nu}$, $\Omega_{\mu\nu}$ is the quantum metric and the Chern number for the ground state(s) defined as 
\begin{align}
    g_{\mu\nu} &= \frac{1}{r}\Tr G^{\mu\nu}, \\
    \Omega_{\mu\nu} &= \frac{1}{r}\Tr F^{\mu\nu}.
\end{align}
If the ground state is unique ($r=1$), $g_{\mu\nu}$ and $\Omega_{\mu\nu}$ reduces to the Abelian quantum geometric tensor. 

If we further assume that the bulk conductivity $\sigma(\omega)$ does not depend on the boundary condition, we can replace $g_{\mu\nu}$ and $\Omega^{\mu\nu}$ with the averaged ones over $\vec{\kappa}$, and in particular $\Omega_{\mu\nu}$ is replaced by $C_{\rm tot}/r$ with the Chern number of the ground state subspace $C_{\rm tot}$.   

With the sum rules~\eqref{eq:SWM_ap}, \eqref{eq:KK_ap}, we can repeat the same discussion in Ref.~\cite{onishi_quantum_2024-1} to show the relation between the structure factor or the quantum weight $K$ and the quantum metric for 2D systems as 
\begin{align}
    K = \int \frac{\dd^2{\vec{\kappa}}}{2\pi}(g_{xx} + g_{yy}),
\end{align}
and this is the generalization of Eq.~(13) in the main text to many-body systems with $U(1)$ symmetry.
From an inequality $g_{xx}+g_{yy} \ge \abs{\Omega_{xy}}$, we can show the bound on the quantum weight $K\ge \abs{C}$ and the topological bound~(11) in the main text.

\section{Topological bound in $d+id$ topological superconductors}
As an example of a topological superconductor, consider a $d_{x^2-y^2}+id_{xy}$ superconductor in two dimensions
~\cite{volovik_fractional_1989}. A model for such a state on a lattice is described by the following mean-field Hamiltonian~\cite{senthil_spin_1999}: 
\begin{align}
    H &= \sum_{\vec{k}} \begin{pmatrix}
        \cre{c}_{\vec{k}\up} & \ani{c}_{-\vec{k}\down}
    \end{pmatrix} \begin{pmatrix}
        \ene(\vec{k}) & \Delta(\vec{k}) \\
        \Delta(-\vec{k})^* & -\ene(\vec{k})
    \end{pmatrix} \begin{pmatrix}
        \ani{c}_{\vec{k}\up} \\
        \cre{c}_{-\vec{k}\down}
    \end{pmatrix}
    \nonumber \\
    &\equiv \sum_{\vec{k}} \cre{\Psi}_{\vec{k}} \qty(\vec{h}(\vec{k})\vdot\vec{\sigma}) \ani{\Psi}_{\vec{k}}.
\end{align}
$\ene(\vec{k})$ is the single particle dispersion and here we take $\ene(\vec{k}) = \ene_0(\cos k_x + \cos k_y)$, and $\Delta(\vec{k})=\Delta_0(\cos k_x - \cos k_y) - i \Delta_{xy} \sin k_x \sin k_y$ is the pairing potential. As in Eq.~(18), we have a topological bound for the $U(1)$ charge $\cre{\Psi}\ani{\Psi}$ and hence for the spin structure factor with $s_0 = \hbar/2$.

This system is in a topological phase with the quantized spin Hall conductance determined by Chern number when $\Delta_{xy}\neq 0$~\cite{senthil_spin_1999}. This can be understood with the Dirac dispersion at four points, $\vec{k}=(\pm \pi/2, \pm \pi/2)$. When $\Delta_{xy}=0$, the quasiparticle excitation at these points has gapless Dirac dispersion, but once $\Delta_{xy}$ becomes finite, these Dirac points are gapped. The Chern number is the sum of the contributions from each Dirac point and is $C=2~\mathrm{sgn}(\Delta_{xy})$. The Chern number can be also directly calculated by 
    $C = (1/4\pi)\int\dd^2{\vec{k}}~ \vec{n}\vdot\qty(\pdv*{\vec{n}}{k_x}\cross\pdv*{\vec{n}}{k_y})$,
with $\vec{n}=\vec{h}/\abs{\vec{h}}$.

We can verify our inequality~(17) by explicitly calculating the spin static structure factor with the Fourier transform of the spin density operator $s_{\vec{q}}^z=(\hbar/2)\sum_{k}(\cre{c}_{\vec{k}\up}\ani{c}_{\vec{k}+\vec{q}\up}-\cre{c}_{\vec{k}\down}\ani{c}_{\vec{k}+\vec{q}\down})=(\hbar/2)\sum_{\vec{k}}\cre{\Psi}_{\vec{k}}\ani{\Psi}_{\vec{k}+\vec{q}}$. The leading order in $q$ of the spin static structure factor is given by $S^s_{\vec{q}}=(\hbar^2/16\pi)q^2 K +\dots$ with 
$K = (1/8\pi)\int\dd^2{\vec{k}}~\sum_{\alpha=x,y}\qty(\pdv*{\vec{n}}{k_\alpha})^2$.
From a mathematical inequality $(\pdv*{\vec{n}}{k_x})^2 + (\pdv*{\vec{n}}{k_y})^2\ge 2\abs{\qty(\pdv*{\vec{n}}{k_x})\cross\qty(\pdv*{\vec{n}}{k_y})}$, one can explicitly confirm the inequality~(17) holds.

\section{Details on calculations of twisted TMD Materials}
The effective Hamiltonian for each valley and spin is given by~\cite{wu_topological_2019}
\begin{align}
    H = \begin{pmatrix}
        -\frac{\hbar^2(\vec{k}-\vec{\kappa}_+)^2}{2m^*} + \Delta_1(\vec{r}) & \Delta_T(\vec{r}) \\
        \Delta_T^{\dagger}(\vec{r}) & -\frac{\hbar^2(\vec{k}-\vec{\kappa}_-)^2}{2m^*} + \Delta_2(\vec{r}) 
    \end{pmatrix}, \label{eq:TMD_moire}
\end{align}
where $\kappa_{\pm} = 2\pi/a_M(-1/\sqrt{3}, \pm 1/3)$
is the corner of the moir\'e Brillouin zone, $a_M = a/\theta$ is the moir\'e lattice constant with the twist angle $\theta$ and the monolayer lattice constant $a$. $\Delta_T(\vec{r})$ and $\Delta_{1,2}(\vec{r})$ are spatially modulated by the moir\'e superlattice of the following form:
\begin{align}
    \Delta_{1,2}(\vec{r}) &= 2V\sum_{j=1,3,5} \cos(\vec{g}_j\vdot\vec{r}\pm\psi), \\
    \Delta_{T}(\vec{r}) &= w(1+e^{-i\vec{g}_2\vdot\vec{r}}+e^{-i\vec{g}_3\vdot\vec{r}}),
\end{align}
where $\vec{g}_1 = (4\pi/\sqrt{3}a_M, 0)$ is the moir\'e reciprocal lattice vector and $\vec{g}_j$ are $(j-1)\pi/3$ counter-clockwise rotation of $\vec{g}_1$. The parameters for twisted WSe$_2$ is given in Ref.~\cite{devakul_magic_2021} as: $m^*=0.43m_0, V=\SI{9.0}{\milli\electronvolt}, \psi=\ang{128}, w=\SI{18}{\milli\electronvolt}, a_0=\SI{3.317}{\angstrom}$ with the bare electron mass $m_0$, while the parameters for MoTe$_2$ is given by~\cite{reddy_fractional_2023} $m^*=0.62m_0, V=\SI{11.2}{\milli\electronvolt}, \psi=\ang{91.0}, w=\SI{13.3}{\milli\electronvolt}, a_0=\SI{3.52}{\angstrom}$.

We calculated the static structure factor from which the Bragg peak divergence is subtracted:
\begin{align}
    S_{\vec{q}}' &= \frac{1}{V}(\expval{\rho_{\vec{q}}\rho_{-\vec{q}}}-\expval{\rho_{\vec{q}}}\expval{\rho_{-\vec{q}}}).
\end{align}
$S_{\vec{q}}'$ reduces to the static structure factor $S_{\vec{q}}$ defined in the main text if $\vec{q}$ does not coincide with the reciprocal lattice vectors. 
$S_{\vec{q}}'$ is calculated by diagonalizing the Hamiltonian~\eqref{eq:TMD_moire} in $k$-space to obtain the Bloch wavefunction for each band and plugging them into the following expression for $S_{\vec{q}}'$:
\begin{align}
    S_{\vec{q}}' &= \int_{\rm BZ}\frac{\dd^d{\vec{k}}}{(2\pi)^d} \Tr[P(\vec{k})(P(\vec{k})-P(\vec{k}+\vec{q}))], \label{eq:BI_Sq_ap}
\end{align}
where $P(\vec{k})=\sum_{n}^{\rm occ} \ket{u_{n\vec{k}}}\bra{u_{n\vec{k}}}$ is the projection operator onto the occupied bands at wavevector $\vec{k}$, and $\ket{u_{n\vec{k}}}$ is the cell-periodic Bloch wavefunction for $n$-th band. The expression~\eqref{eq:BI_Sq_ap} was derived in, for example, Supplemental Materials of Ref.~\cite{onishi_quantum_2024-1}, but we give the derivation of ~\eqref{eq:BI_Sq_ap} in the next section for the sake of completeness. In the diagonalization, we consider plane waves with wavevector of the form $\vec{k}=\vec{k}_0 + n_1 \vec{g}_1 + n_3 \vec{g}_3$, where $\vec{k}_0$ is the wavevector in the first Brillouin zone and $n_{1, 3}$ is integer between $-3$ and $3$. The integration over the Brillouin zone is approximated by the summation over $30\times 30$ mesh in the first Brillouin zone. In our calculation, $P(\vec{k})$ is for the occupied bands of the holes, and for filling factor $\nu=-1$, $P(\vec{k})$ is the projection operator onto the highest valence band. 

We note that $S_{\vec{q}}'$ is not periodic in $\vec{q}$ even for periodic systems. For example, $S_{\vec{q}}'$ at finite reciprocal lattice vector $\vec{G}$ is not equal to $S_{\vec{q}=0}$; the former represents the density fluctuation with wavevector $\vec{G}$ and finite in general, while the latter represents the density fluctuation with wavevector $\vec{q}=0$, which is zero in charge conserved systems. 

The quantum weight is calculated by integrating the quantum metric:
\begin{align}
    K &= \int_{\rm BZ}\frac{\dd^2{\vec{k}}}{2\pi} (g_{xx}(\vec{k}) + g_{yy}(\vec{k})). \label{eq:K-BI_ap}
\end{align}
where the quantum metric is calculated with the following formula:
\begin{align}
    g_{\mu\nu}(\vec{k}) &= \Re\sum_{n}^{\rm occ}\sum_{m}^{\rm unocc}\frac{\mel{u_{n\vec{k}}}{\partial_\mu H}{u_{m\vec{k}}}\mel{u_{m\vec{k}}}{\partial_\nu H}{u_{n\vec{k}}}}{(\ene_{n\vec{k}}-\ene_{m\vec{k}})^2}
\end{align}
where $\ene_{n\vec{k}}$ is the band dispersion for $n$-th band, $\partial_\mu$ is the derivative with respect to $k_{\mu}$, and the summation of $n$ (m) is over the occupied (unoccupied) bands. The integration in Eq.~\eqref{eq:K-BI_ap} is executed with a code of Julia by hcubature function in HCubature package with a relative tolerance of \SI{1e-4}{}.

\begin{figure*}
    \centering
    \includegraphics[width=1.0\columnwidth]{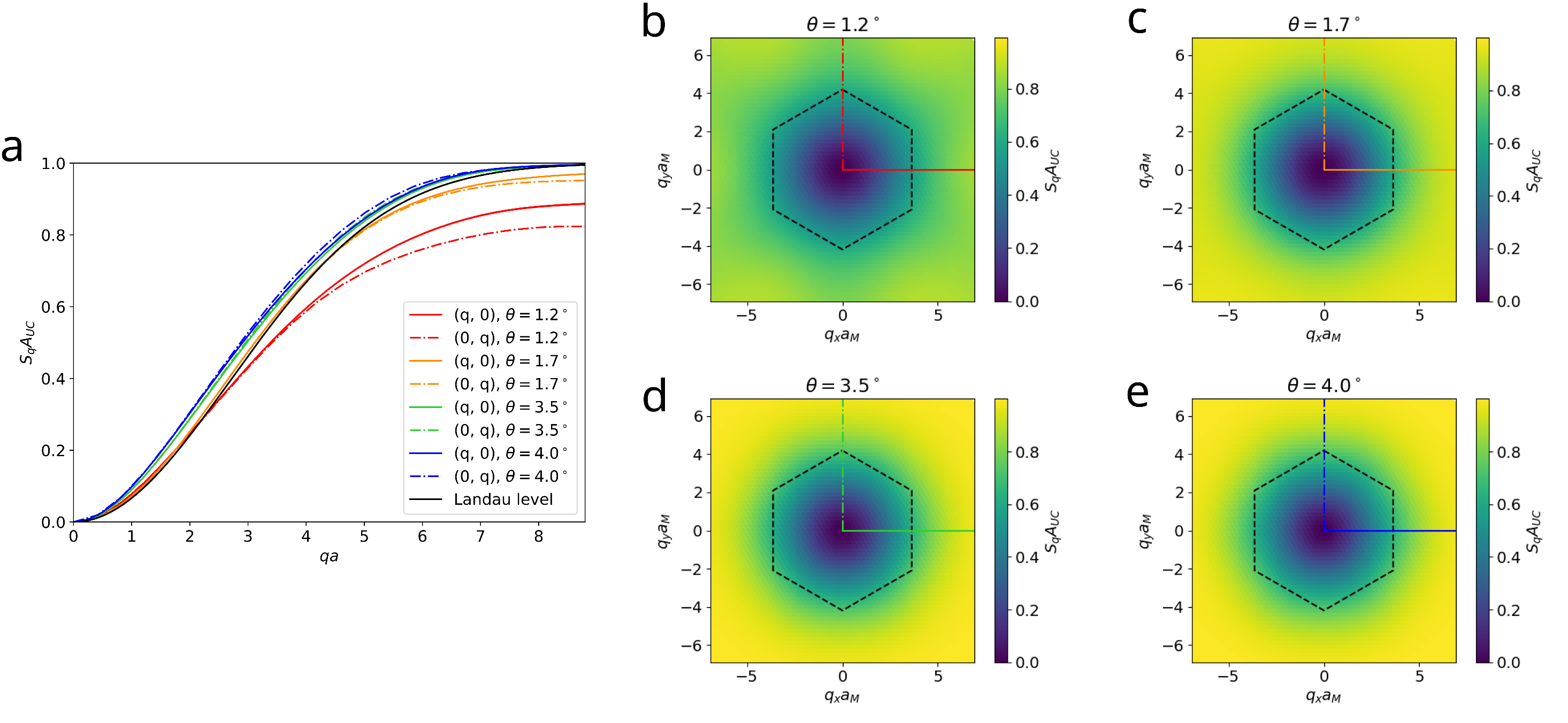}
    \caption{Structure factor of twisted MoTe$_2$.}
    \label{fig:SM_MoTe2}
\end{figure*}
Results for other twist angles are shown in Fig.~\ref{fig:SM_MoTe2}.

\section{Feynmann-Bijl formula for general systems with $U(1)$ symmetry }

Here we derive the Feynman-Bijl formula for general $U(1)$ symmetric systems. To find an upper bound on the energy gap, we consider constructing an ansatz for an excited state $\ket{ex}$ so that $\ket{ex}$ is orthogonal to the ground state $\ket{gs}$. As long as $\ket{ex}$ and $\ket{gs}$ are orthogonal to each other, the energy expectation value of $\ket{ex}$ is always larger than or equal to the first excited state. Therefore, we can find an upper bound on the energy gap by evaluating the energy difference between $\ket{ex}$ and $\ket{gs}$. We choose the following state as $\ket{ex}$:
\begin{align}
    \ket{ex} &= (\rho_{-\vec{q}}-\expval{\rho_{-\vec{q}}}_{gs})\ket{gs},
\end{align}
where $\expval{\dots}_{gs}$ is the expectation value in the ground state $\ket{gs}$ and $\rho_{\vec{q}}$ is the $U(1)$ charge density. The orthogonality $\ip{ex}{gs}=0$ is easily confirmed. Now, the energy gap $\Delta$ has an upper bound: $\Delta \le \mel{ex}{H}{ex}/\ip{ex}{ex}-\expval{H}_{gs}$. This can be rewritten as 
\begin{align}
    \Delta \le \frac{\expval{\comm{\rho_{\vec{q}}}{\comm{H}{\rho_{-\vec{q}}}}}_{gs}}{2VS_{\vec{q}}}, \label{eq:general_FB_ap}
\end{align}
with the structure factor $S_{\vec{q}}$ for the $U(1)$ charge in the ground state. 

For electronic systems, $U(1)$ charge $\rho$ can represent the electric charge and the Hamiltonian $H$ is quadratic in the momentum $\vec{p}$ with kinetic energy $\vec{p}^2/(2m)$ for each electron. Using a relation from the continuity equation, $\comm{H}{\rho_{-\vec{q}}} = -\hbar\vec{q}\vdot\vec{j}_{-\vec{q}}$, and the commutation relation $\comm{\rho_{\vec{q}}}{\vec{j}_{-\vec{q}}}=\hbar \vec{q}N/m$ with the number of electron $N$ and the mass $m$, the numerator is reduced to a simple form and the gap bound is obtained as 
\begin{align}
    \Delta \le \frac{\hbar^2\vec{q}^2 n}{2m S_{\vec{q}}}. \label{eq:Feynman-Bijl_ap}
\end{align}
This is the Feynman-Bijl formula used in Ref.~\cite{feynman_atomic_1954}. The same formula as ~\eqref{eq:general_FB_ap}, but with the projected density operator $\bar{\rho}_{\vec{q}}$ and the projected Hamiltonian $\bar{H}$ instead of the full density operator $\rho_{\vec{q}}$ and the full Hamiltonian $H$, is used in a study of magneto-roton excitations in fractional quantum Hall systems~\cite{girvin_magneto-roton_1986}.

\section{Structure factor of noninteracting electronic systems}
The static structure factor of noninteracting electronic systems (Eq.~(12) in the main text) was calculated in, for example, Ref.~\cite{onishi_quantum_2024-1}. However, we derive the expression here for the sake of completeness.    
Writing the cell-periodic Bloch wavefunction for $n$th band as $\ket{u_{n\vec{k}}}$ with the wavevector $\vec{k}$, the number density operator with wavevector $\vec{q}$ is given by 
\begin{align}
    n_{\vec{q}} &= \sum_{n, m} \sum_{\vec{k}} \ip{u_{n,\vec{k}}}{u_{m,\vec{k}+\vec{q}}} \cre{c}_{n{\vec{k}}}\ani{c}_{m,\vec{k}+\vec{q}},
\end{align}
with $\ani{c}_{n,\vec{k}}$ the annihilation (creation) operator of electron in $n$th band with wavevector $\vec{k}$. Then the static structure factor for finite $\vec{q}$ is given by 
\begin{align}
    S'_{\vec{q}}&= \frac{1}{V}\sum_{n,m,\vec{k}}\sum_{n',m',\vec{k}'}\ip{u_{n,\vec{k}}}{u_{m,\vec{k}+\vec{q}}}\ip{u_{n',\vec{k}'}}{u_{m',\vec{k}'+\vec{q}'}}\qty(\expval{\cre{c}_{n{\vec{k}}}\ani{c}_{m,\vec{k}+\vec{q}}  \cre{c}_{n'{\vec{k}'}}\ani{c}_{m',\vec{k}'+\vec{q}'}}-\expval{\cre{c}_{n{\vec{k}}}\ani{c}_{m,\vec{k}+\vec{q}}}\expval{\cre{c}_{n'{\vec{k}'}}\ani{c}_{m',\vec{k}'+\vec{q}'}}) \\
    &= \frac{1}{V}\sum_{n,m,\vec{k}}\ip{u_{n,\vec{k}}}{u_{m,\vec{k}+\vec{q}}}\ip{u_{m,\vec{k}+\vec{q}}}{u_{n,\vec{k}}}(1-f_m(\vec{k}+\vec{q}))f_n(\vec{k})
\end{align}
where $f_n(\vec{k}) = \expval{\cre{c}_{n\vec{k}}\ani{c}_{n\vec{k}}}$ is the occupation number of $n$th band at wavevector $\vec{k}$. Defining the projection operator $P(\vec{k}) = \sum_{n}^{\rm occ} \ket{u_{n\vec{k}}}f_n(\vec{k})\bra{u_{n\vec{k}}}$, we obtain an expression for $S'_{\vec{q}}$ for a general noninteracting system as 
\begin{align}
    S'_{\vec{q}}&= \frac{1}{V}\sum_{\vec{k}}\Tr[P(\vec{k})(1-P(\vec{k}+\vec{q}))] = \frac{1}{V}\sum_{\vec{k}}\Tr[P(\vec{k})(P(\vec{k})-P(\vec{k}+\vec{q}))]
\end{align}
where we have used $P(\vec{k})^2 = P(\vec{k})$ in the last equality. Further rewriting the summation over $\vec{k}$ with integral, we obtain:
\begin{align}
    S'_{\vec{q}}&= \int_{\rm BZ}\frac{\dd^d\vec{k}}{(2\pi)^d}\Tr[P(\vec{k})(P(\vec{k})-P(\vec{k}+\vec{q}))]. \label{eq:Sq_BI_P}
\end{align}
In particular, for band insulators, the projection operator becomes $P(\vec{k})=\sum_{n}^{\rm occ} \ket{u_{n\vec{k}}}\bra{u_{n\vec{k}}}$ and Eq.~\eqref{eq:Sq_BI_P} reduces to Eq.~\eqref{eq:BI_Sq_ap} above. Noting that $S'_{\vec{q}}=S_{\vec{q}}$ except when $\vec{q}$ coincides a reciprocal lattice vector in lattice systems, Eq.~(12) in the main text also follows from Eq.~\eqref{eq:Sq_BI_P}.

\end{widetext}

\end{document}